# Revolutionary Hybrid E-Books for Enhanced Higher Learning

Dr. RD.Balaji,
Salalah College of Applied Sciences,
Sultanate of Oman,
balajicas@yahoo.com

Dr. Ramkumar Lakshminarayanan,
Sur College of Applied Sciences,
Sultanate of Oman,
rajaramcomputers@gmail.com

Er. Malathi Balaji,
Madurai Kamaraj University,
India,
bmalathisai@gmail.com

## ABSTRACT

Books are the best friends of human beings and make them a rational animal. In this E-world our traditional books are losing their values. Recent surveys are proving that the younger generations are not much interested in visiting libraries and reading books due to their addiction towards the electronic gadgets. Many publishers are now changed their strategy by publishing and promoting e-books. Academicians are also badly affected by this trend and they are forced to motivate the students to improve their reading habits for their better performance in the institutions and to be responsible citizens of a country.

There are many difficulties faced by the e-book readers and it discourages the people to read an e-book for long time like a normal printed book. The main objective of this paper is to introduce a hybrid E-book which is capable of having audio and video files also in it. This makes the people to read the e-book with modern electronic formats. This multimedia facility makes the people to read or listen or watch the e-books for long time. We have quoted the survey results which prove that the students prefer to read hybrid books than normal e-books or the printed handouts in College of Applied Sciences, Oman. Still we feel these books should have few more facilities, so that people who like traditional books will also adopt these e-books, without losing the satisfaction of reading printed books. These things are listed in the recommendations and conclusion section.

## INTRODUCTION

In these days, the researchers and the common people come across many multi disciplinary researches. The educational researchers mostly belong to interdisciplinary studies. If we see the definition of education, it is the process of transferring the knowledge, skills, culture etc., via teaching from one generation to the next via learning [2]. Learning is the process of having a change in the mental or behavioral state of a person because of some processes like reading, watching and much more. The use of computers in education has mostly been focused on enhancing teaching and learning in formal settings, typically in the traditional classrooms and labs [10]. Instead of making the technology as a supporting tool for the traditional teaching, nowadays new era of educational field is being started which moved from traditional education model to totally technology based Massive Open Online Courses (MOOCs) education. This is supported by the flipped classroom concept, e-books and e/m-learning technology [11].

## E/M LEARNING

"*Any type of learning that takes place in learning environments and spaces that take account of the mobility of technology, mobility of learners and mobility of learning*" [3]. The above definition is given for m-learning and it is mostly accepted by many of the researchers. M-learning is also viewed as a component of e-learning. M-learning started emerging after 2004 rapidly because of the advancements in the mobile phone devices and tabs introduction. Like any other technologies m-learning also was having lot of issues in the infant stage and by now it has overcome most of the disadvantages, as well as new facilities also provided to the m-learners which was not expected even by researchers in the initial stages of m-learning [5]. The Cloud concepts in the IT field influenced most of the m-learning strategies in providing common materials in a most economical way internationally with security. Many researchers have compared the m-learning with e-learning and agreed that m-learning is the extension of e-learning and e-learning with anywhere-anytime facility becomes m-learning [9]. In a recent survey with the higher education students in Oman, proved that 100 percent of the students are having smart phone and they are well equipped with the basic infrastructure required for implementing the m-learning. Hence we have decided to start doing our research with the m-learning environment and also provided options to the students to go with e-learning. In the recent mobile phones we are having more interactive options to make communication with a group of people easily [4]. So we have decided to go for m-learning environment than e-learning because of the above mentioned advantages.

## FLIPPED CLASSROOM

The researchers found that revising the missed class or revising the taught concept is the major time waster for the academicians. The lecturers make use of the multimedia technologies to share the notes and ideas what they have taught to the students in the class room, so that students may revise or learn the topics apart from their class hours with the help of the recorded classes. This is called flipped classroom. There are many recent technologies like QR Code, E-books used by the lecturers to create flipped classrooms to utilize the student's time even after their class hour for teaching [7] [8]. "*The flipped classroom inverts traditional teaching methods, delivering instruction online outside of class and moving "homework" in to the class room*" [6]. The flipped classroom concept helps average and below average students to learn their course at their own pace in an effective way. There are many social websites, learning management tools and other contemporary chat tools help learners to interact with each other and with lecturers at anytime and anywhere. This helps to overcome the e-learning drawback of not having face to face interaction with teachers even after their class timings. In the flipped classroom the main concept used is the activity based learning [6]. The flipped classroom concept was started from 2007 by the academicians. Most of the MOOCs courses are using this concept and running successfully. Many researchers proved that the flipped classroom helps to improve the understanding of the students in a specific topic and give better result in overall [7] [8]. Flipped classrooms are very much helpful to the students to revisit the lectures when they don't understand the concept, since the lectures are small in size (8 to 15 minutes). We are trying to use the same concept in the Colleges of Applied Sciences for a specific course in the form of hybrid e-book so that students will get more time and comfort for the revision on concepts and programming skill practices.

## E-BOOK CREATING TOOLS

There are so many e-book creation software's available in the market for either nominal cost or for free. Each and every tool has its own advantages and disadvantages. The comparison of these tools is showed in table 1. We can expect more options from these tools in near future. The stakeholder's expectations from these tools are discussed in the recommendation section.

| Brand Name / Feature | Open Source | Page Flip | e-Archive | Audio/Video | Software Installation | Publish/Share Online | Work Online | Work Offline | For PC | For Mac |
|---|---|---|---|---|---|---|---|---|---|---|
| flipb Software | No | Yes | Yes | Yes | Yes | Yes | Yes | Yes | Yes | Yes |
| ePaperFlip | No | Yes | Yes | Yes | No | Yes | Yes | No | Yes | Yes |
| PUB HTML5 | No | Yes | Yes | Yes | Yes | Yes | Yes | Yes | Yes | Yes |
| kvisoft flipbook maker pro | No | Yes | Yes | Yes | Yes | Yes | Yes | Yes | Yes | Yes |
| 3D Issue | No | Yes | Yes | Yes | Yes | Yes | Yes | Yes | Yes | Yes |
| Slide HTML5 | No | Yes | Yes | Yes | Yes | Yes | Yes | Yes | Yes | Yes |

| | | | | | | | | | |
|---|---|---|---|---|---|---|---|---|---|
| Adobe Photoshop | No | No | No | No | Yes | No | No | Yes | |

Table 1: e-book creation software comparison (Source: http://en.wikipedia.org/wiki/List_of_E-book_software)

## IMPLEMENTATION

To perform the implementation, we selected two higher educational institutions in Oman for implementing the hybrid e-book as a course material for the students.  We took much effort in the sample selection for this research so that the outcome would be accurate. We were concentrating on the equality in the samples selected. We had selected one group from each college where the average performance of the students was same in the pre-requisite course. We kept the usual materials in the Blackboard and gave links to the video lectures that was available in the YouTube through Blackboard itself. We were having same procedures to conduct the normal classes and exams. Same question papers were used to evaluate the student's performance in this course. We asked one college students to install the hybrid e-book. Another college was not provided with the hybrid e-book. Using the statistical option given in Blackboard we were monitoring the number of students accessing the Blackboard during their semester in both the colleges. Also using Student Information System (SIS) software's result options, we were comparing the results and dropout numbers of the students from both the colleges. This research was conducted for one semester. At the end of the semester the collected data were analyzed and produced in the next section. Apart from the collected data, students were asked to provide feedback about the hybrid e-book and asked them to give periodic report of the usage of e-book and its difficulties in using it. An academic team was employed in preparing e-book with required content and short video of the lectures and the lab exercise demonstration. We were very specific about making the content as short as possible to make it easy and comfortable for the students to read. For all other interactive sessions we had taken the help of Blackboard. In general we named the college as A which distributes the hybrid e-book to students and another one as college B.

## ANALYSIS

The research team desired to know the impact of hybrid e-book in the accessibility of the Blackboard. So the research team was focusing on the number of times the Blackboard was accessed for viewing the materials and also number of posts in the forum. The table 2 is clearly showing that the students of college A have not accessed the materials more than 2 times on an average throughout the semester. But the college B students have accessed the materials on an average of 8 times. The forum access was more from the college A students than college B. They were having more than 24 posts on an average by each student compared to college B students, who were having an average of 11 posts throughout the semester.

| | College A | College B |
|---|---|---|
| **Accessed Course Materials** | 2 | 8 |
| **Accessed Forum for discussions** | 24 | 11 |

Table 2: Accessibility of Blackboard by college A and B students

In table 3 we have recorded the registration and result details of college A and B students. College A students who received the hybrid e-book did not withdrawn the course in the middle, where in college B, 2 students left in the mid of the course. Similarly all the students passed the subject from college A , where 2 students failed the course from college B.

| | College A | College B |
|---|---|---|
| **No. of students registered** | 25 | 25 |
| **No. of students withdrawn from course in the middle** | 0 | 2 |
| **No of students passed the course** | 25 | 21 |

Table 3: Students registration and result details of college A and B

From the above two tables we can easily find out that the college A students performance is better compared to college B. since both the college students are having similar background and handled by the same research team, we strongly feel that hybrid e-book motivates the college A students and college B students are lacking.

In the below given graphs 1 and 2, let us see the grade distribution among college A and college B students. These graphs also again prove that the grades of college A students are better than the college B.

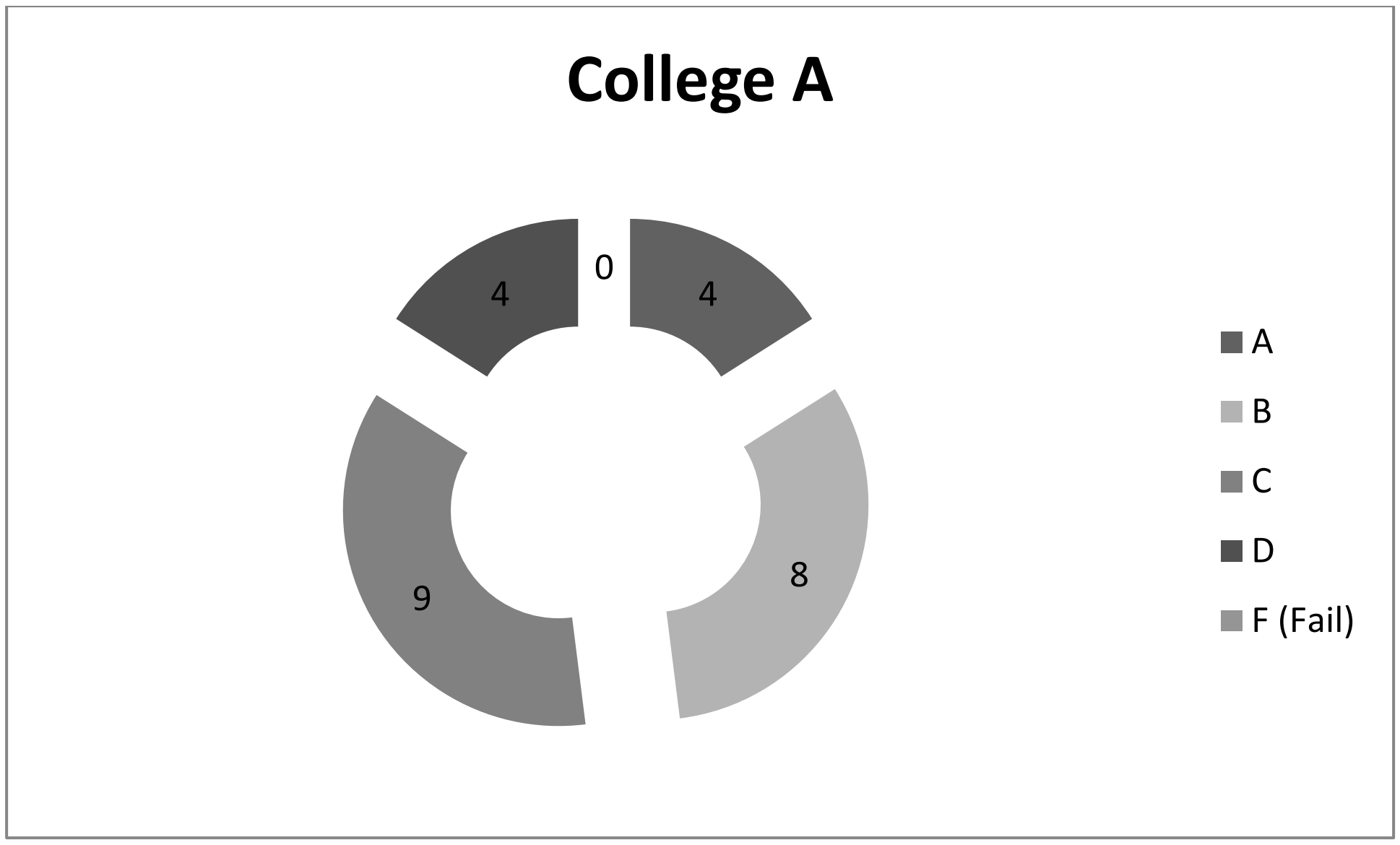

Graph 1: Grade distribution of College A students

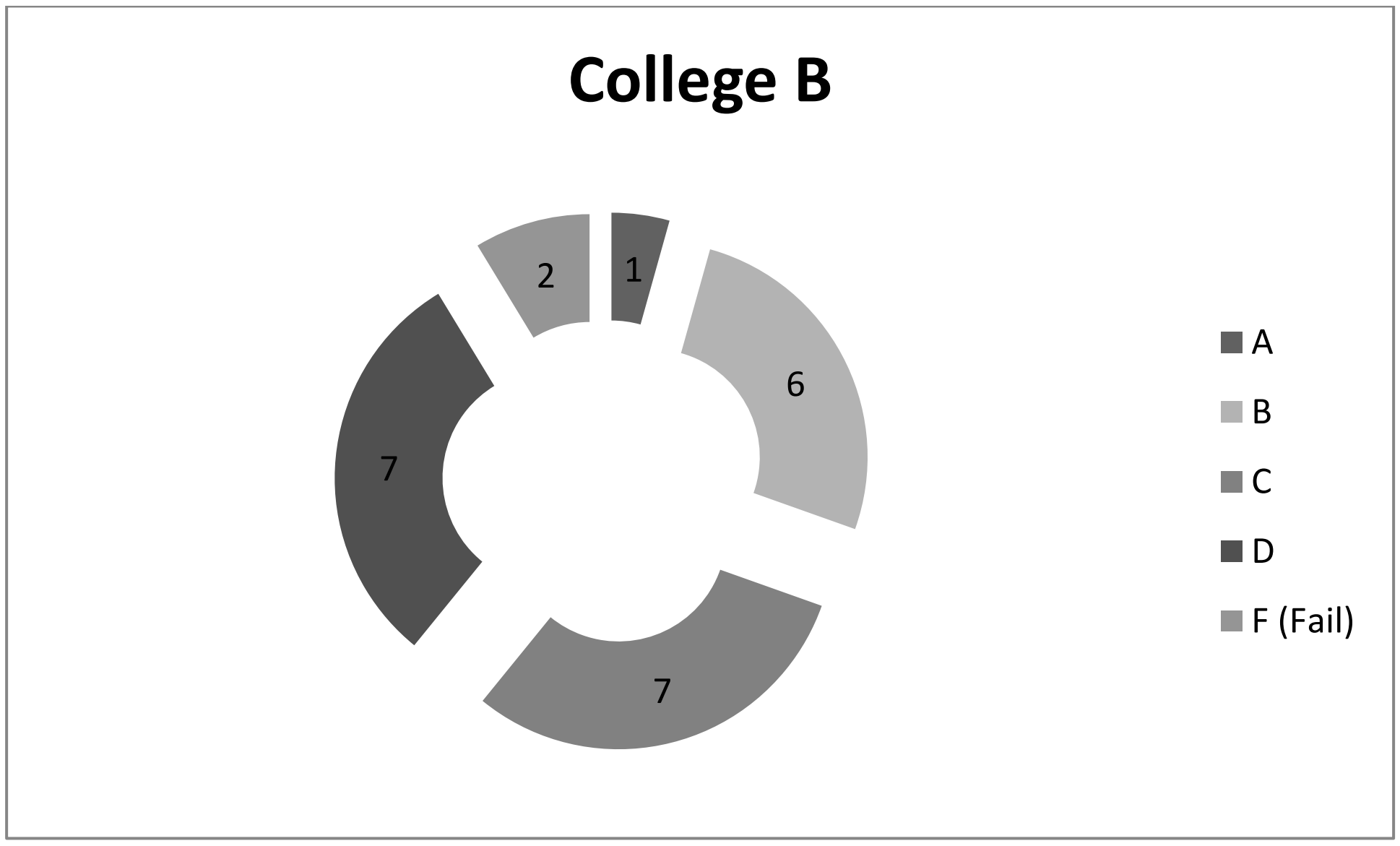

Graph 2: Grade distribution of College B students

**RECOMMENDATION**

In the previous section we have discussed the advantages of hybrid e-books in the higher education institutions in the Sultanate of Oman. Still we have found that the students are having few difficulties in using hybrid e-books and expecting few more options for the maximum benefit from the hybrid e-books. The hybrid e-books are prepared in English and few students are having difficulty in understanding the language, they want to have dictionary option in the hybrid e-book itself. In majority of the hybrid e-books “Note Taking” options are missing. Students are willing to have the note taking option in their mother tongue or it should help the students to accept handwritten option. Hybrid E-books are having lots of advantages, but when a student wants to interact with teacher or wants to have access to the forum for that course, there is no facility provided. When a student accesses the hybrid e-books online, they want to have the forum facility also to be accessed. Due to more multimedia facilities in the hybrid e-books, the accessibility of the books is comparatively slow. Hence the speed of accessing the books should be fast compared to the accessibility speed now. Since this research is conducted in the limited number of higher education institutions in Oman, we have got only few suggestions and able to identify limited drawbacks with the hybrid e-books. By expanding this research to other higher educational institutions we can improvise the hybrid e-books to a great extent.

## CONCLUSION

Most of the world universities, colleges and other academic institutions are working towards teaching and learning which can make the students to access the study materials by JIT (Just-In-Time) and anywhere-anytime. This will help the students to access the materials at their potential time and reduces the knowledge loss. This paper is discussing about the hybrid e-books implementation at the higher education institutions in the Sultanate of Oman and also proves that this methodology helps the students to perform well in their exams and other academic activities. This can be achieved by the students when they have hazel free method to access the materials relevant to their course. M-learning is the excellent tool to act as a platform for the implementation of hybrid e-books. Even though M-learning is solving all the issues in the implementation of hybrid e-books, still we keep the e-learning option open to the students who wish to use this sometimes. This paper also proves that students prefer M-learning than E-learning for the access of these hybrid e-books. Similarly hybrid e-books help students to get in touch with the materials as and when they wish, and improve their performance in their exams. Still there are few disadvantages in the current version of hybrid e-books which are pointed out in this paper. By improving these options in future we can make these revolutionary e-books as a flipped classroom tool and also they can be used in MOOCs for the improved educational environment with minimum knowledge loss.